\documentclass[prl,twocolumn,floatfix,amsmath,amssymb]{revtex4}
\usepackage{graphicx,color}% Include figure files
\usepackage{dcolumn}% Align table columns on decimal point
\usepackage{bm}% bold math
\begin{document}

\title{Spin-liquid versus dimerized ground states in a frustrated Heisenberg antiferromagnet}
\author{Luca Capriotti,$^{1}$ Douglas J. Scalapino,$^{2}$ and Steven R. White$^{3}$}
\affiliation{
${^1}$ Kavli Institute for Theoretical Physics, University of California, Santa Barbara, California 93106-4030 \\
${^2}$ Department of Physics, University of California, Santa Barbara, California 93106-9530  \\
${^3}$ Department of Physics and Astronomy, University of California, Irvine,
California 92697
}

\date{\today}

\begin{abstract}
We present a Density Matrix Renormalization Group (DMRG)
study of the ground-state properties of spin-1/2
frustrated $J_1-J_3$ Heisenberg $n_l$-leg ladders (with $n_l$ up to 8).
%For low values of the $J_3/J_1$ ratio, 
%we find results consistent with long-range antiferromagnetic order in
%the two-dimensional thermodynamic limit. 
%In contrast, in the regime of stronger
%frustration ($J_3/J_1\simeq 0.5$) our results indicate that the 
%two-dimensional system is a homogeneous spin liquid with a finite spin gap.
%only even-leg ladders display a 
%finite gap to spin excitations. This gap decreases as the 
%numbers of legs increases, consistently with long-range 
%antiferromagnetic order in the two-dimensional limit.
%corresponding 
%to a gapless phase with long-range antiferromagnetic order 
%in the two-dimensional limit. 
For strong
frustration ($J_3/J_1\simeq 0.5$), both even- and odd-leg 
ladders display a finite gap to spin excitations,
which we argue remains finite in the two-dimensional limit.
In this regime, on odd-leg ladders the ground state
is spontaneously dimerized, in agreement with the Lieb-Schultz-Mattis
prediction, while 
on even-leg ladders the dimer correlations decay exponentially.
The magnitude of the dimer order
parameter decreases as the number of legs increases, consistent with a 
two-dimensional spin-liquid ground state. 
%We support a recently proposed scenario where the two-dimensional limit 
%can still be consistently defined in this situation
%as the odd-leg dimer order parameter vanishes for infinite number 
%of chains, thus leading to a spin-liquid ground state in two-dimensions.
\end{abstract}

\vspace{3cm}

%%%%%%%%%%%%%%%%%%
%%%%%%%%% PACS USED
%%%%  71.10.Fd Lattice fermion models (Hubbard model, etc.)
%%%%  74.20.Mn Nonconventional mechanisms (spin fluctuations, polarons and bipolarons, resonating valence bond model, anyon mechanism, marginal Fermi liquid, Luttinger liquid,
%%%%  71.10.Pm Fermions in reduced dimensions
%%%%  71.27.+a Strongly correlated electron systems; heavy fermions
%%%%%%%%%%%%%%%%%%

\maketitle

Despite many years of intense investigations, 
the existence of a homogeneous spin-liquid ground state 
for a spin-1/2 system on a two-dimensional square lattice
remains controversial.
This is mainly because there has been no definite evidence so far 
that a microscopic model could stabilize a homogeneous 
non-magnetic phase with one electron per unit cell.  
In fact, the known non-magnetic phases of spin-1/2
quantum antiferromagnets 
in one or two dimensions usually display a spontaneously broken 
translation symmetry related to spin-Peierls dimerization, as 
in the frustrated Heisenberg chain \cite{mygosh,j1j2chain} and
ring exchange models, \cite{ring}, or arising from a doubling of the 
ground-state unit cell, as in the Heisenberg two-leg ladder.\cite{twoleg} 

For one-dimensional systems a rigorous result, the 
Lieb-Schultz-Mattis (LSM) theorem, \cite{lsm} implies that a gapped 
non-magnetic phase is in general associated with
a broken translation symmetry. This result can be also
extended to spin-1/2 Heisenberg models defined on
odd-leg ladder geometries.\cite{affleck} 
%Clearly, to understand to what extent
%the LSM result can be applied to two-dimensional systems, it is
%of crucial importance to ascertain the existence or non-existence
%of homogeneous spin-liquid phases on the square lattice. 
There have been several recent attempts to generalize the LSM result
to two dimensions \cite{bonesteel,oshikawa,hastings}. Here it has been argued
that the gapped phase is associated with a ground-state degeneracy. However, 
there are different opinions on whether this degeneracy necessarily
implies a spontaneously broken translation symmetry (and thus the 
non-existence of a two-dimensional spin-liquid) or whether it is associated
with a topological degeneracy of fractionalized spin-liquid phases
\cite{bonesteel,wen,lhuillier,ivanov}.
Recently,  on the basis of a variational approach, 
Sorella {\em et al.},\cite{chiral} 
have proposed that a spin-liquid ground state can be stabilized 
in two-dimensions and  yet satisfy the constraint imposed by the LSM theorem.
In fact, within the formalism of projected BCS wave functions
it is possible to construct a gapped state which displays
spontaneous dimerization on any odd-leg ladder
thus satisfying the LSM theorem, but with no dimerization
for even-leg ladders. The two-dimensional thermodynamic limit 
is consistently reached for a large number of legs since the dimer 
order parameter on odd-leg ladders vanishes in this limit, 
thus leading to a homogeneous
spin liquid. In this case, therefore, the ground-state 
degeneracy predicted by the 
generalizations of the LSM theorem \cite{hastings} is not 
connected to a spontaneously broken translation symmetry 
but rather to a topological degeneracy of fractionalized
resonating valence bond states.\cite{ivanov,chiral}

In this paper, we examine  $n_l$-leg 
frustrated Heisenberg ladders with Hamiltonian
\begin{equation} \label{ham}
\hat{H}=J{_1}\sum_{\langle i,j\rangle}
\hat{{\bf {S}}}_{i} \cdot \hat{{\bf {S}}}_{j}
+ J{_3}\sum_{\langle\langle i,j\rangle\rangle}
\hat{{\bf {S}}}_{i} \cdot \hat{{\bf {S}}}_{j}~~.
\end{equation}
Here ${\bf \hat{S}}_{i}$ are spin-$1/2$ operators on a square lattice, and
$J_1,J_3\ge 0$ are the nearest- and third-nearest neighbor antiferromagnetic
couplings along the two coordinate axes.
In the following, we use the numerical density matrix renormalization
group (DMRG) \cite{dmrg} to study the ground state of this Hamiltonian on ladder 
systems with $n_l$ legs of length $L$ with open boundary conditions. 
In our calculations, we typically performed 15-20 sweeps of the lattice, keeping
a maximum of $m \simeq 2000$ states and obtaining discarded weights 
smaller than $\sim 5 \times 10^{-7}$. 
Our plan is to carry out DMRG calculations for ladders with different 
number $n_l$
of legs. Then, by extrapolating in the length $L$ of the ladders and
looking at the behavior of the odd-and even leg systems for modest value 
of $n_l$ we seek to gain insight into the behavior of the two-dimensional
system.

The classical ground state of the $J_1{-}J_3$ Hamiltonian
in two dimensions displays conventional N\'eel order for 
$J_3/J_1\le 0.25$.
For  $J_3/J_1 > 0.25$  the ground state has 
incommensurate antiferromagnetic order with a 
pitch vector depending on the frustration ratio, assuming 
the value $Q=(2\pi/3,2\pi/3)$ 
at $J_3/J_1=0.5$, and approaching $Q=(\pi/2,\pi/2)$,
corresponding to four decoupled N\`eel lattices,
for $J_3/J_1\to \infty$.
For the case of quantum spin-1/2, in two-dimensions, 
the ground state is expected
to display long-range N\'eel order for $J_3/J_1 \to 0$,
and numerical calculations on lattices up to 32 sites suggest
that a non-magnetic ground state could be stabilized in the 
regime of strong frustration $J_3/J_1\sim 0.5$.\cite{leung}
This work also found signatures of dimerization for 
values of $J_3/J_1 \simeq 0.7$ in agreement with the predictions
of series expansions.\cite{singh}

\begin{figure}
\vspace{-20mm}
\includegraphics[width=0.43\textwidth]{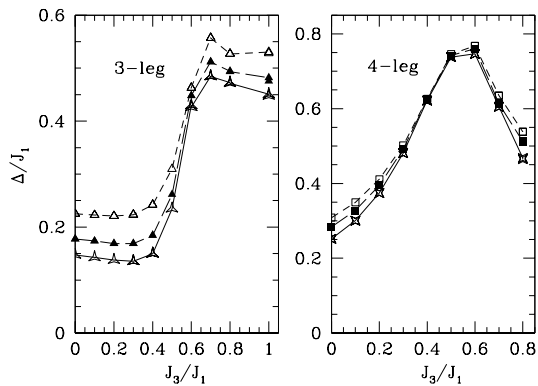}
\vspace{-10mm}
\caption{\label{gapvsj}
Spin gap as a function of the ratio $J_3/J_1$. Left: 3-leg ladder for
$N=12$ (empty triangles), 16 (full triangles), and 20 (stars). Right: 4-leg
ladder for $N=10$ (empty squares), 12 (full squares), and 16 (stars).
}
\end{figure}

The effects of frustration on the antiferromagnetic correlations of the
ground state can be investigated by studying the 
behavior of the spin-gap as a function of $J_3/J_1$.
In fact, in a system with long-range spin correlations, spin excitations are 
necessarily gapless. Instead, if the energy cost of the lowest 
triplet excitation, $\Delta$, remains finite in the thermodynamic limit 
the ground state has short-range spin-correlations.
As shown in Fig.~\ref{gapvsj}, for $n_l=3$ and 4, the spin gap  
increases as the frustration ratio $J_3/J_1$ increases. This is seen
for both the even- and the odd-leg ladder case. In particular, in the odd-leg 
case, which is gapless with power-law spin correlations 
in the pure Heisenberg limit,\cite{noack} the spin gap due to the
finite length of the ladder 
remains almost constant for small values 
of $J_3/J_1$ but increases sharply for $J_3/J_1\simeq 0.4$, 
suggesting a transition to a gapped non-magnetic phase.
Alternatively, in the even-leg ladder case, a finite correlation length
is expected for small $J_3/J_1$ and
the spin-gap increases smoothly with $J_3/J_1$ as
no magnetic transition is expected.
In both cases, the spin gap reaches a maximum for 
intermediate values of $J_3/J_1$ where the effects of frustration are 
expected to be the strongest, then it decreases again for large $J_3/J_1$
when the limit of four decoupled Heisenberg lattices is eventually recovered.

\begin{figure}
\includegraphics[width=0.41\textwidth]{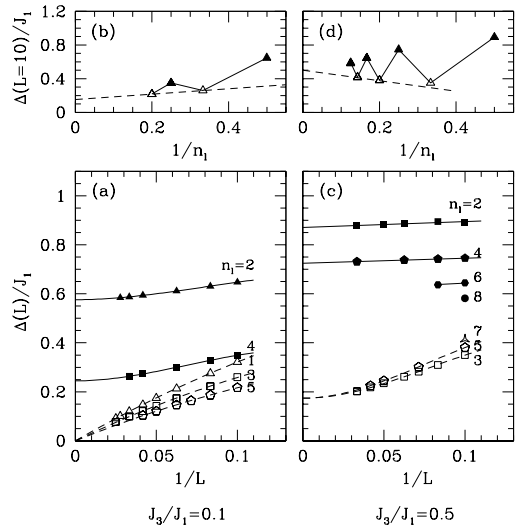}
\vspace{-1mm}
\caption{\label{ssgap}
Size scaling of the spin gap for $J_3/J_1=0.1$ [(a) and (b)], and $J_3/J_1=0.5$
[(c) and (d)]. (a) and (c): spin gap as a function of the length of the
ladders, $L$, for different number of legs, $n_l$. (b) and (d): spin gap
as a function of $n_l$, for $L=10$. Empty (full) symbols correspond to
odd- (even-) leg ladders. Lines are guides for the eye.
}
\end{figure}

The size scaling of the spin gap is shown in Fig.~\ref{ssgap}.
For weak frustration, the spin gap extrapolates 
to zero for the odd-leg ladders, and 
to a constant, which decreases with $n_l$, for the even-leg ladders.\cite{sudip}
This is consistent with the gapless
N\`eel ordered phase expected in the two-dimensional limit.
Instead, for $J_3/J_1=0.5$ the spin gap extrapolates 
to a constant as $L\to \infty$ for both the even- and the odd-ladders
we have studied. The difference between the regime of low and high frustration 
is also seen from the dependence of the spin-gap 
on the number of legs 
for a fixed chain length $L$ (see also Fig.~\ref{ssgap}-(b) and (d)). 
For low-frustration the spin-gap decreases with the number of legs 
both for even (full symbols) and odd (empty symbols) $n_l$. 
However, in the regime of high frustration
it decreases with $n_l$ only for even leg samples while it {\em increases} 
with the number of legs on odd-leg samples. This behavior
is consistent with a two-dimensional
phase which has exponentially decaying spin correlations.

The presence of a finite-gap in the excitation spectrum
of the $n_l$-leg ladders has consequences in view of the LSM theorem. 
In fact, on
odd-leg ladder systems it is possible to construct an excitation
in the singlet sector with momentum $(\pi,0)$ which
becomes degenerate with the ground state in the thermodynamic limit.\cite{lsm}
This implies either a gapless spectrum or, in the presence of a finite gap,
a two-fold degenerate ground state with a doubling of the unit cell 
and a spontaneously broken translation symmetry. In the one dimensional model
this is known to be realized through spin-Peierls dimerization.\cite{mygosh}
Instead, the LSM result does not apply to even-leg ladders
so that, in these geometries, both translationally invariant and dimerized ground states are 
in principle compatible with a finite triplet gap. 

\begin{figure}
\includegraphics[width=0.41\textwidth]{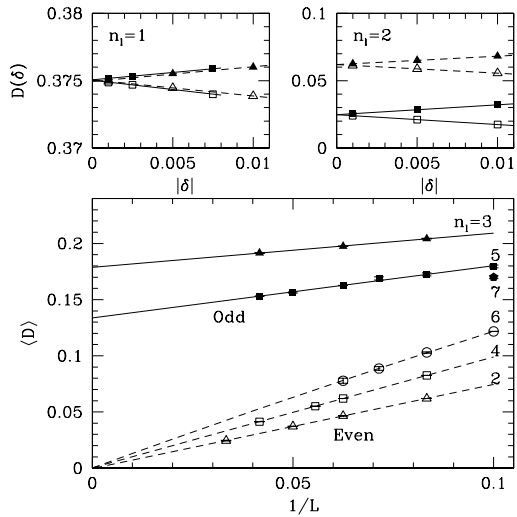}
\vspace{-5mm}
\caption{\label{dimer}
Upper panels: $D(\delta)$ for $J_3/J_1=0.5$. Left: one dimensional
(Majumdar-Gosh) chain for $N=12$ (triangles), and 24 (squares).
Right: 2-leg ladder for $N=12$ (triangles), and 30 (squares). Full (empty)
symbols correspond to $\delta<0$ ($\delta>0$). Lower panel:
size-scaling of the dimer order parameter, $D(\delta\to0)$, as a function
of the length of the ladders, $L$, for different numbers of legs, $n_l$.
}
\end{figure}

The occurrence of spin-Peierls dimerization can be studied by
calculating the response of the system to a nearest-neighbor spin-spin 
operator which breaks the translation symmetry along the chains with 
momentum $Q=(\pi,0)$
\begin{equation}\label{dimerpar}
\hat{O}=\sum_{r} e^{i Q\cdot r}\hat{{\bf {S}}}_{r} \cdot \hat{{\bf {S}}}_{r+x}~.
\end{equation}
Here $x=(1,0)$ is a unit vector along the chain direction.
This can be done in general by adding to the Hamiltonian
a term $\delta \hat{O}$ where $\delta$ is a (small) generalized field.
The order parameter can be calculated as the limit for $\delta\to 0$ of
the ground-state expectation value of $\hat{O}$ in presence of the field,
$D=\lim_{\delta\to 0} \langle \hat{O}\rangle_\delta/N$,
where $N=L\times n_l$ is the total number of sites. 
On periodic finite-size systems $D$ vanishes in general by symmetry 
as it breaks the translational invariance and the symmetry under 
{\em site-centered} lattice reflections along the chain direction of 
the unperturbed Hamiltonian (\ref{ham}).
However, on samples with an even length $L$ and open boundary conditions, 
the dimer order parameter (\ref{dimerpar}) will be in general non-zero and 
can be calculated using the Hellmann-Feynman theorem as
$D=de(\delta)/d\delta |_{\delta=0}$.
%
%
%This definition, however, is usually of little utility for numerical 
%studies which are performed on periodic finite-size systems, where 
%$D$ vanishes in general by symmetry as it breaks the translational 
%invariance and the symmetry under {\em site-centered} lattice reflections
%along the leg direction of the unperturbed Hamiltonian.
%Instead, on samples with an even length $L$
%and open boundary conditions, the dimer order parameter (\ref{dimerpar}) 
%will be in general non-zero and can be calculated using the
%Hellmann-Feynman theorem as
%$D=de(\delta)/d\delta |_{\delta=0}$.
%
%
%However, since the DMRG method uses open boundary conditions, one has 
%the possibility of a direct calculation of the order parameter 
%on a finite-size system. In fact, in a system with open boundaries the 
%momentum is not a definite quantum number of the Hamiltonian and the 
%only symmetries that can be broken by the spin-Peierls operator on a ladder 
%geometry are {\em site centered} lattice reflections along the leg direction, under 
%which $\hat{O}\to -\hat{O}$. However, on a finite-size ladder,
%these reflections are symmetries of the Hamiltonian only for odd 
%$L$. Hence, on samples with an even length $L$
%and open boundary conditions, the spin-Peierls order parameter will
%be in general non-zero and can be calculated using the 
%Hellmann-Feynman theorem as
%$D=de(\delta)/d\delta |_{\delta=0}$. 
Here $e(\delta)$ is the ground-state energy
per site (in unit of $J_1$) in the presence of the perturbation. 
As a result, within the DMRG technique
the dimer order parameter can be calculated with simple energy 
measurements by computing  $e(\delta)$ for a few values of $\delta$ 
and then estimating numerically the limit
$D=\lim_{\delta\to 0} (e(\delta)-e_0)/\delta$. This is illustrated
in the upper panels of Fig.~\ref{dimer} for a single chain
and a two-leg ladder at $J_3/J_1=0.5$.
Here, as a consistency check, the dimer order parameter 
is estimated by calculating the limit for $\delta\to0$ of 
$D(\delta)=(e(\delta)-e_0)/\delta$ both
for positive (filled symbols) and negative (empty symbols) 
$\delta$'s. The two limits converging to the same value.
In particular, for the one-dimensional chain at the exactly solvable point
$J_3/J_1=0.5$ (Majumdar-Gosh model) the known size-independent result,
$D=0.375$, is recovered \cite{mygosh}.

The size scaling of the dimer order parameter obtained with this procedure
is shown in the same figure for ladders with different numbers of legs. 
This analysis reveals close similarities with the variational 
scenario of Ref.~\cite{chiral}.
In fact, in the odd-leg ladder case the dimer order parameter extrapolates to a constant for infinite chain length, as required by the LSM theorem.
Instead, in the even-leg cases, where the system is not constrained by the 
LSM theorem to dimerize, the dimer order parameter extrapolates 
to zero. 

\begin{figure}
\includegraphics[width=0.41\textwidth]{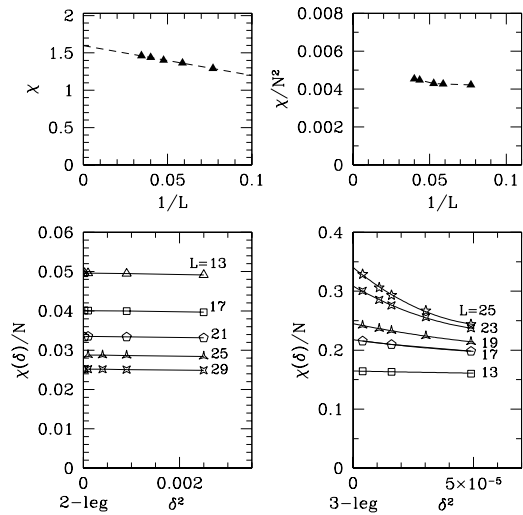}
\vspace{-5mm}
\caption{\label{susk}
Dimer susceptibilities for the 2-leg (left panels) and 3-leg (right panels)
ladders. The lower panels show $\chi(\delta)/N$ {\em vs}
$\delta$ for various lengths,
and the top panels the size scaling of the extrapolated values,
$\chi=\lim_{\delta\to 0}\chi(\delta)$ (see text). Note the different normalizations
for the 2- and 3-leg cases in the upper panels. 
Lines are guides for the eye.
}
\end{figure}

This conclusion is supported also by the calculation of 
dimer susceptibilities.  These can be calculated within our numerical approach
by considering ladders with odd length $L$. In these geometries the 
unperturbed Hamiltonian is reflection symmetric around the 
central column of sites, so that 
the order parameter $D$ vanishes by symmetry
for any finite-size cluster, and the ground-state energy 
has corrections proportional to $\delta^2$.
Therefore $e(\delta)\simeq e_0-\chi \delta^2 /2$ , with
$\chi$ the generalized susceptibility associated with the operator
$\hat{O}$, namely,
$\chi = 2 \langle \psi_{0}| \hat{O} (E_{0}-\hat{H})^{-1}
\hat{O} | \psi_{0} \rangle/N$.
If true long-range order in the dimer correlations exists in the
thermodynamic ground state, the finite-size susceptibility
will diverge as the system size increases.
In particular, it can be shown that it is bounded from below by the
system volume squared, $\chi \sim N^2$.~\cite{chibound} 
Thus susceptibilities are a sensitive
tool for detecting the occurrence of long-range order. 
In analogy with the calculation of the order parameter, the susceptibility
$\chi = -d^2e(\delta)/d\delta^2|_{\delta=0}$
can be calculated numerically from
$
\chi= \lim_{\delta \to 0} \chi(\delta)=-{2(e(\delta)-e_0)}/{\delta^2}~,
$
as illustrated in the bottom panels of Fig.~\ref{susk} for 2- and 
3-leg ladders.

The behavior of the susceptibilities
for the even- and odd-leg ladders is remarkably different (Fig.~\ref{susk}). 
In particular,
$\chi/N$ decreases with the linear size $L$ in the even-leg ladder case, and 
it increases with $L$ in the odd-leg ladder case. 
The susceptibilities  for the odd-leg ladders appear 
to diverge as $N^2$, as required for true long-range dimer order.
In contrast, for the even-leg ladders the 
susceptibilities are bounded, indicating
short-range dimer correlations with a finite correlation length.

\begin{figure}
\includegraphics[width=0.45\textwidth]{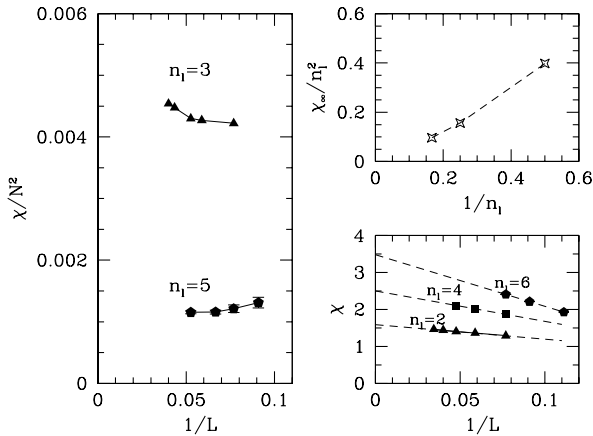}
\vspace{-18mm}
\caption{\label{fig5}
Left: Size scaling of $\chi/N^2$ for the 3- and 5-leg ladders as a
function of the chain-length $L$.
Right bottom: Size scaling of $\chi$ for the 2-, 4- and 6-leg ladders
as a function of the chain-length $L$ (bottom). Right top: Size scaling of the
$L\to\infty$  extrapolated values of the even-leg ladder susceptibilities,
$\chi_\infty$, as a function of the numbers of legs, $n_l$.
}
\end{figure}

In order for the two-dimensional limit to exist, the ground-state correlations
for even and odd-leg ladders must converge in the limit of 
a large number of legs. This is the case, for instance, in the limit of 
small frustration of this model, where 
the spin correlations decay exponentially for even $n_l$ and with a power-law
for odd $n_l$. Here the two-dimensional limit is reached as
the correlation length for the even-leg ladders diverges exponentially 
with $n_l$ leading to long-range antiferromagnetic order
in two dimensions. \cite{noack,sudip} 
As we have shown, in the regime of strong frustration, the dimer-correlations
are short-ranged on even-leg ladders while a finite-dimer order parameter
is observed on odd-length ladders. 
However, as the number of chains $n_l$ is increased, 
the odd-leg ladder dimer order parameter decreases 
(Fig.~\ref{dimer}), and the divergence of the dimer susceptibility becomes 
weaker (see left panel of  Fig.~\ref{fig5}). 
On the other hand, the infinite-$L$ dimer susceptibility, 
$\chi_\infty$, on even-leg ladders does not appear
to diverge as the square of the number of chains, $n_l^2$,
(see right-panels of Fig.~\ref{fig5}) as one would expect in presence
of long-range dimer order in two dimensions.
Thus, a ground-state with no spontaneous broken translation symmetry 
in the two-dimensional limit appears as a plausible 
interpretation of our results.

In conclusion, we have shown that a spin-gapped ground state with short-range antiferromagnetic correlations
is stabilized by frustration on the spin-1/2 $J_1{-}J_3$ model 
at $J_3/J_1=0.5$ on ladders with $n_l=$ 1 to 8 legs. The behavior of the
spin gap by increasing the number of legs is consistent with a non-magnetic
ground state in the two-dimensional limit.
On odd-leg ladders we find a finite dimer order parameter associated with
a spontaneously broken translation symmetry. However, as the number of legs increases $n_l=1,3,5,7$ the size of the order parameter decreases and the
divergence of the associated susceptibility becomes weaker.
These results suggests that in the two-dimensional
limit the dimer order parameter vanishes. Although the numerical data
we have presented here was for $J_3/J_1=0.5$, we find similar results
for other values of $J_3/J_1$ near 0.5 and believe that for a range of 
$J_3/J_1$ values this model exhibits a non-magnetic ground state 
in two dimensions.
Our results are consistent with a recently proposed scenario 
\cite{chiral} where 
the odd-leg dimer order vanishes for infinite number of legs 
leading to a homogeneous spin-liquid in two dimensions. 

The authors would particularly like to acknowledge Federico Becca,
Alberto Parola and Sandro Sorella for their contribution to this work
and Matthew Fisher, Duncan Haldane, Michael Hermele, Arun Paramatekanki, and 
Didier Poilblanc for many useful discussions.
This work was supported by NSF under Grants No. DMR02-11166 and  DMR03-11843.

\end{document}